\documentclass[letterpaper,12pt]{article}
\usepackage{jheppub1} % modified JHEP-author-manual

\usepackage[T1]{fontenc} % if needed
\usepackage{soul}
\usepackage{amssymb}
\usepackage{amsmath}
\usepackage{amsfonts}
\usepackage{graphicx}
\usepackage{subfigure}
\usepackage{color}
\usepackage{dcolumn}
\usepackage{hyperref}
\usepackage{comment}
\numberwithin{equation}{section}

\newcommand{\be}{\begin{equation}}
\newcommand{\ee}{\end{equation}}
\def\bea{\begin{eqnarray}}
\def\eea{\end{eqnarray}}
\newcommand{\eq}[1]{(\ref{#1})}
\def\nn{\nonumber}

\newcommand{\beq}{\begin{equation}}
\newcommand{\eeq}{\end{equation}}
\newcommand{\ben}{\begin{eqnarray}}
\newcommand{\een}{\end{eqnarray}}
\newcommand{\bes}{\begin{subequations}}
\newcommand{\ees}{\end{subequations}}
\newcommand{\blg}{\begin{align}}
\newcommand{\elg}{\end{align}}

\newcommand{\prt}[1]{{\left( {#1} \right)}}

\newcommand{\la}[1]{\label{#1}}
\newcommand{\startappendix}{
\setcounter{section}{0}
\renewcommand{\thesection}{\Alph{section}}}

\newcommand{\Appendix}[1]{
\refstepcounter{section}
\begin{flushleft}
{\large\bf Appendix \thesection: #1}
\end{flushleft}}

\def\one{\mbox{1 \kern-.59em {\rm l}}}

\def\a{\alpha}

\def\k{\kappa}
\def\l{\lambda} 
\def\m{\mu} \def\n{\nu}

\def\s{\sigma}  
\def\t{\tau}

  \def\cF{{\cal F}}

  \def\cO{{\cal O}}

  \def\cX{{\cal X}}

\def\where{{\quad\mbox{where}\quad}}
\begin{document}

\hfill{WITS-CTP-154}

\vspace{15pt}

\begin{center}

{\Large \bf
Tension of Confining Strings at Low Temperature}
%\vspace{35pt}
\vskip 0.5in

{\bf
 Dimitrios Giataganas$^{1,2}$\phantom{x}and\phantom{x} Kevin Goldstein$^{3}$
}
\vskip 0.2in
{\em
{}$^1$Department of Nuclear and Particle Physics,\\
Faculty of Physics, University of Athens,\\
Athens 15784, Greece
\vskip .15in
{}$^2$ Rudolf Peierls Centre for Theoretical Physics,\\
University of Oxford, 1 Keble Road,\\
Oxford OX1 3NP, United Kingdom\vskip .15in
{}$^3$ National Institute for Theoretical Physics,\\
School of Physics and Centre for Theoretical Physics,\\
University of the Witwatersrand,\\
Wits, 2050, South Africa
}
\vskip .15in
{\small \sffamily
dgiataganas@phys.uoa.gr, kevin.goldstein@wits.ac.za
}

\vspace{90pt}
{\bf Abstract}\end{center}

In the low temperature confining phase of QCD or QCD-like theories it
is challenging to capture the temperature dependence of observables
through AdS/CFT. Using the blackfold approach we compute the
quark--anti-quark linear static potential in the low temperature confining phase, taking into account the thermal excitations of the string.
We find the explicit temperature dependence of the string tension and notice that, as  naturally expected, tension decreases as temperature increases. We have also generalized the blackfold approach for the computation of the Wilson loops, making it directly applicable to a large class of backgrounds.

\setcounter{page}0
\newpage

\section{Introduction}

The static potential between a heavy quark and anti-quark pair,
generated by a thin gluonic flux-tube formed between the pair, is of
fundamental interest. Confining theories at zero temperature are characterized by a
linearly increasing potential at large separation distances
with the proportionality constant given by the string tension. The linear
behavior in the potential comes from the world-sheet of the
string while its quantum fluctuations produce the universal subleading
L\"{u}scher term. At low temperatures, below the deconfinement
transition temperature, string models predict a decrease of the
potential -- in particular the string tension decreases as
temperature increases. For large separation lengths it has been
found that the string tension, $\sigma$, goes like
$\s=\s_0-c T^2$, where $\s_0$ is the zero temperature string tension, $T$ is the temperature and $c$ is a
constant \cite{Pisarski:1982cn,deForcrand:1984cz,Gao:1989kg}. For
a certain range of parameters these predictions were confirmed with
Monte Carlo lattice data \cite{Bakry:2010zt}.

In addition to the string tension, the temperature affects other
quantitative and qualitative properties of the confined Q\={Q}
state. For instance, the flux tube profile differs from the zero
temperature one. At zero temperature, it has been found using a string
model \cite{Luscher19811} and confirmed by several lattice
calculations \cite{Caselle:1995fh}, that the width broadens
logarithmically as a function of the bound state size. At low
temperature this broadening is modified \cite{Bakry:2010zt},
increasingly, deviating from the zero temperature result as the temperature increases. In fact, when one
approaches the deconfinement temperature, the broadening becomes
linear \cite{Allais:2008bk}.

There have been extensive efforts using string theory
models and lattice calculations to understand the temperature dependence of the binding dynamics of a
heavy Q\={Q} pair.  In our paper, we extend these investigations
using gauge/gravity duality \cite{adscft1,adscft2}. We calculate
the string tension of a heavy bound state in a
confining theory at low temperature, taking into account the thermal
excitations of the string.

To our knowledge this is the first top-down  AdS/CFT computation
that captures the temperature dependence of the string tension. The
challenge in setting up this computation is that the low temperature solutions of  confining
backgrounds do not have a black hole horizon. One can increase
the temperature but this leads to a Hawking-Page transition and to
a deconfined theory with a black hole. This is a first order transition
since there is no smooth solution connecting the two phases. Alternatively, there is a simple geometrical reason to
explain the absence of a black hole in the low temperature case.
Black hole horizons, associated with a temperature, and  cigar type geometries, associated with confinement, are generated by two qualitatively compactifications -- the former is along the Euclideanised time direction
 while the latter is space-like . When the time
circle shrinks to zero, the spatial circle does not and vice
versa. In other words, one either has a black
hole horizon or a scale that introduces  confinement encoded in the metric. This is a
reason that in confining gauge/gravity theories it is difficult to study the temperature dependence of certain observables. A way to overcome this
problem was to introduce bottom-up models to
phenomenologically study the static potential
\cite{Andreev:2006ct,Ghoroku:2005kg,BoschiFilho:2006pe}. In our paper
we capture part of the temperature dependence of the bound state
quantities holographically using an alternative more natural method,
the blackfold approach, where the string world-sheet corresponding to
the Wilson loop, is modified by the thermal excitations generated by
the finite temperature of the theory.

We find a temperature dependent string tension, $\s=\s_0-c T^3/N$
where $N$ is the number of colors in the theory and $c$ is a known
constant, depending on the number of Q\={Q} pairs of the bound state. The decrease in the string tension is in qualitative
agreement with string model predictions and physical
expectation. Note that we did not necessarily expect our result to
match with the $T^n$ scaling of the string tension for the single quark pair found using effective string models given the differences between the two bound states, the techniques used and the physics that they probe, but the results might be complementary. We elaborate further on this in our paper.

The blackfold approach is an effective theory which aims to describe black
branes whose worldvolume is not in stationary equilibrium or flat. It
is valid in the probe approximation, for small deviations from the flat
stationary black branes on scales much longer than the brane thickness
\cite{Emparan:2009cs,Emparan:2009at,Emparan:2011br,Emparan:2011hg}. For
our purpose we use  blackfolds to capture the thermal
excitations of the string world-sheet corresponding to a Wilson loop
in a heat bath. The probe black string is placed in
a finite temperature background in thermal equilibrium. This method
has already  been used  for  Wilson loops in  finite temperature
$AdS$ space \cite{Grignani:2012iw}, in Schr\"{o}dinger space with
scaling $z=2$ \cite{Armas:2014nea} and  giant gravitons in $AdS$
\cite{Grignani:2013ewa}, as well as in other cases, for example
\cite{Grignani:2010xm,Niarchos:2012cy}.

We find  generic equations, obtained from applying the blackfold
approach to black string probes dual to a Wilson loop, which can
be applied directly to  appropriate backgrounds\footnote{Our results do not apply for backgrounds which necessarily have off diagonal terms in the metric which would occur if, for example, one has rotation.}. It is
interesting that this generalization is possible, and allows a direct
substitution of the metric elements to obtain the energy of the string
in particular backgrounds. This happens with several observables
that can be expressed in terms of metric elements using the Nambu-Goto
action, including the Wilson loop expectation value
\cite{Kinar:1998vq,Giataganas:2012zy,Giataganas:2013lga,Brandhuber:1999jr}.

To calculate the expectation value of the Wilson loop, rather than the
Nambu-Goto action \cite{Maldacena:1998im}, we use the free energy of
the probe, which in the zero temperature limit reproduces the
Nambu-Goto result.  We use $k$ separate, fundamental strings with
$1\ll k\ll N$ and $g_s^2 L\gg 1$. The separate $k$ strings should not
be confused with
$k$-strings. The $k$-strings correspond to Wilson loops in higher symmetric and
antisymmetric representations, and in the  dual gravity  side are
represented as $D3$ and $D5$ branes respectively with electric fluxes
\cite{Gomis:2006sb,Gomis:2006im,Yamaguchi:2006tq}. In the blackfold
approach a more plausible explanation is that we have $k$ separate
Q\={Q} flux tubes, that  do not
interact at zero temperature. Increasing the
temperature we find the energy of the $k$ separate strings in thermal
equilibrium with the background. We observe that the thermal terms in the
energy of the strings do depend on the number of strings in a way that can not
be factorized as a common factor in the expression. This means that
while bringing the separate strings into thermal equilibrium with the
background, the strings do interact although this interaction may be indirect. We point out that this interaction is different from the
$k$-string bound state computed by the DBI action of the $D3$ or $D5$
branes.

We apply our generic formulas to confining theories. We choose the AdS
soliton background \cite{Horowitz:1998ha}, and we calculate the new
linear static potential taking into account  thermal
fluctuations and we find a decrease in the string
tension. Our work opens new directions, especially in determining the
temperature dependence qualitatively in observables in confining
theories at low temperature.  We remark that in the top-down constructions
this has not been achieved until now.

Our paper is organized as follows. In section 2 we provide generic
techniques for applying the blackfold method to the Wilson loop string
world-sheets for any appropriate background. We obtain the free energy
and the constraints on the validity of the results in terms of the background metric
elements. Having developed the generic formalism we apply it to the
confining $AdS$ soliton background in section 3. We compute the string
tension and we find it reduced compared to the zero temperature
case. We comment on the modified term in the low temperature case as
well as in the high temperature one. We conclude our paper by summarizing
and discussing our results in section 4.

\section{$k$ F-Strings in Blackfold Approach}

In this section, which generalizes \cite{Grignani:2012iw}, we use the
blackfold approach to study the orthogonal Q\={Q} Wilson loop in a
generic background.  In the strong coupling regime we use a number of
probe strings ($k$ of them) to construct the finite temperature dual
of Wilson loops. The general approach we follow below is for a
background with metric $g_{\m\n}$ where the probe extends in two
dimensions for configurations that satisfy the following equilibrium
and boundary conditions \cite{Emparan:2011hg} \be T^{a b} K_{a
  b}^{c}=0~,\quad T^{a b} n_{b}|_\partial=0~,\quad J^{a b c}
n_c|_\partial=0~.  \ee The equilibrium conditions are on the effective
stress energy tensor $T^{ab}$ of the probes and the extrinsic curvature
$K^{c}_{ab}$. The boundary conditions are projected with the use of the
orthogonal covector at the boundary of the space and also include the
case of a charged black probe with an effective current $J$.

By considering the equation of state of the effective fluid that lives
in its worldvolume, we get the effective stress energy tensor of the fluid
depending on the temperature of the theory $T$. However, as we will see below, it
is more convenient to express the temperature dependence of the stress energy tensor implicitly
in terms of other parameters.

In order to examine the probes, we consider the generic background
\be\label{metric1}
ds^2=g_{00}dt^2+g_{ii}d{x^i}^2+g_{uu}du^2+d\cX^2
\ee
in which the probes are placed. The index $i$ is summed over traverse
spatial dimensions and $u$ is the radial coordinate. At the boundary
of the background, $u_b$, the metric elements $g_{ii}$ diverge, so
$g_{ii}(u_b)=\infty$. When the background includes a black hole, its
horizon $u_h$, is located at a zero of $g_{00}$, so $g_{00}(u_h)=0$. $\cX$ is the internal
space of the theory and in our analysis plays no
role  since we consider the string localized at a point on $\cX$. In the most
generic treatment an extension of the string in the internal space is
allowed, but the form of contributions from the external space in the
blackfold approach is not expected to change, apart from certain
constants. This is because the additional terms in the string energy
due to the blackfold approach are generated from the excitations of
the string close to the black hole horizon. Moreover, the additional
form fluxes of the generic background do not play any role in the
following analysis since they do not couple to the string. Only the $B$-field
may play a role, and it couples to the string but its contributions
also do not expected to modify qualitatively the form of the energy
corrections.

To simplify the form of various equations  we define the following quantities
\bea
\label{defs}
f_1(u):=|g_{00}(u)| g_{11}(u)~,\quad f_u(u):=|g_{00}(u)| g_{uu}(u)~,\quad
f_{1u}(u):=\frac{ g_{11}(u)}{g_{uu}(u)}~,
\eea
which will be used in the paper.

%Eventually the scale of the black hole horizon position $u_h$ will be traded with another scale, the temperature $T$, since this is the physical parameter that appears in the field theory. The temperature for a generic background it is derived in the Appendix A and reads ??
%\be
%T=\frac{1}{4 \pi\sqrt{f_u(u)}}\frac{g_{00}}{u-u_h}\bigg|_{u=u_h}
%\ee
To examine the Wilson loop,  we choose the radial gauge configuration for the world-sheet
\be
t=\t,\quad u=\s,\quad x^1=x(\s)~,
\ee
with the conditions at the boundary, $u_b$, of the space being
\be
x(u_b)=0~,\quad x(u_b)=L~.
\ee
The induced metric,  which contains all the background information we need for our analysis, then reads
\be\label{inmetric}
ds^2 = g_{00} d\t^2+(g_{uu}+x'^2g_{11})d\s^2~,
\ee
 where $x':=d x/d\s$.

The solution of $k$-coincident black string probes in IIB supergravity gives the effective temperature, $T_{loc}$ , string tension, $T_s$, and entropy density, $s$, as \cite{Emparan:2011hg}
\be\label{temp1}
T_{loc}=\frac{n}{4\pi r_0\cosh\a}~ ,\qquad  T_{s}=\frac{n A r_0^n\cosh\a \sinh\a}{k}~,\qquad s=4\pi A r_0^{n+1}\cosh\a~,
\ee
where $n:=D-p-3$, $p$ is the black probe spacial dimension, $D$ is the total number of the space-time dimensions,
$r_0$ is a length scale associated with the horizon, $\a$ is a dimensionless charge parameter and $A=\frac{\Omega_{n+1}}{16 \pi G}$.
The energy momentum tensor elements corresponding to energy density and pressure are
\be\label{tens}
\epsilon=T_{00}=A r_0^n(n+1+n\sinh^2 \a),\quad P=T_{11}=-A r_0^n(1+n \sinh^2\a) ~.
\ee
For strings we consider here, $p=1$, so that for ten dimensional supergravity $n=6$.
In the blackfold approach the first law of thermodynamics is equivalent to the equations of motion and therefore the free energy can be used as the action principle. In a generic background with a redshift factor $R_0$, the free energy for the string probe (keeping the charge $k$ fixed) reads
\be\la{freeen1}
\cF=\int dV R_0 \left(\epsilon-T_{loc}s\right)=A\int dV R_0 r_0^6(1+6 \sinh^2\a)~,
\ee
where $V$ is the spacial volume of the world-sheet.
Using (\ref{temp1}) to eliminate $Ar_0^6$ we can also write the free energy as
\begin{equation}
  \label{eq:freeenTS}
  \cF=kT_s\int dV R_0 \left(\frac{1+6 \sinh^2\a}{6\cosh\a\sinh\a}\right)~.
\end{equation}
 The redshift factor $R_0$ and the local temperature measured from an asymptotic observer in the induced metric \eq{inmetric} are
\be
\label{Tloc}
T_{loc}=\frac{T}{R_0}~,\qquad R_0(\s)=\sqrt{|g_{00}|}~.
\ee
Therefore to achieve asymptotic equilibrium of the probe with  the background we need to equate the local temperature with the temperature \eq{temp1} obtaining a relation between $\a$ and $r_0$:
\be\label{r0}
r_0=\frac{3\sqrt{|g_{00}|}}{2\pi T\cosh \a}~.
\ee
Having specified all the quantities appearing in \eq{freeen1} we  proceed to solve the resulting equations of motion.

\subsection{Free Energy of the Q\={Q} configuration}

In this section we derive the static potential from the blackfold
approach for generic supergravity finite temperature backgrounds. Using the definitions (\ref{defs}), the form of the induced metric and the equation (\ref{r0}), the free
energy (\ref{freeen1}) in the radial gauge for the thermal $F$-string probe in a generic background reads
\be \cF=A \left(\frac{3}{2\pi T}\right)^6\int d\s
\sqrt{f_u(u)}\sqrt{1+ f_{1u}(u)
  x'^2}|g_{00}|^3\frac{1+6\sinh^2\a}{\cosh^6\a}~.
\ee
Varying the free energy with respect to $x(\s)$ leads to the equation of motion
\be\label{xp} x'=\left(\frac{ f_{1u}^2 G(\s)^2}{c^2}-
  f_{1u}\right)^{-1/2}~,
\ee
where $c$ is an integration constant and
$G(\s)$ is defined as
\be
G(\s):=\sqrt{f_u}~|g_{00}|^3\frac{1+6\sinh^2\a}{\cosh^6\a}~.%\quad\mbox{where}\quad B_1(u):=g_{00}^3\sqrt{g_{00}g_{uu}}
\ee
The constant $c$, can be related to the turning point of the
string $x(\s_0)$ (where $x'(\s_0)=0$). It is easily seen from
(\ref{xp}) that
\be c^2= f_{1u}(\s_0) G(\s_0)^2~.
\ee
Due to symmetry, the turning point of the string configuration is located at
$x(\s_0)=L/2$. Therefore, the distance $L$ between the Q\={Q} pairs is
obtained by \eq{xp} and is equal to
\be\label{l1}
\frac{L}{2}=\int_{\s_b}^{\s_0}d\s\left(\frac{ f_{1u}^2(\s) G^2(\s)}{
    f_{1u}(\s_0)G^2(\s_0)}-f_{1u}(\s)\right)^{-1/2}~.
\ee
Now using (\ref{eq:freeenTS}) and $T_s=1/\prt{2\pi l_s^2}=\sqrt{\lambda}/\prt{2\pi R^2}$ (with $\lambda$ the t'Hooft parameter and $R$ the $AdS$ radius),
it is convenient to write the free energy as
\be \cF=\frac{\sqrt{\l} k}{\pi
  R^2}\int_{\s_b}^{\s_0} d\s \sqrt{f_u}\sqrt{1+ f_{1u} x'^2}Y~,
\ee
with
\be
Y=\tanh \a+\frac{1}{6 \cosh \a \sinh \a}~,
\ee
and $\s_b$ is the
boundary of the space either at $u\rightarrow \infty$ or $u\rightarrow
0$. To avoid possible sign errors we choose the
boundary at $\s_b=0$ without loss of generality. The results can be
easily converted to the case $\s_b=\infty$ where $\s_b>\s_0$.

\subsection{Length scales and constraints}\label{sec:con}

The validity of the blackfold approach for  Wilson loop computations
puts strong constraints on the parameters of the model which
subsequently constrain the turning point of the world-sheet and the
length $L$ of the string.  The thickness of the F-string $r_c=r_0
(\cosh\a \sinh\a)^{1/6}$ should be smaller than the other basic length
scales in the problem. In a generic case with
asymptotic $AdS$, we require that $r_c$ is much smaller than the $AdS$
radius ie. $r_c\ll R$. In most of the cases another scale is
introduced in the radial direction -- such as a black hole horizon or
a natural cut off of the space (coming for example from a spatial
compactification and leading to cigar type geometries)\footnote{For
  example, in the presence of a black hole, the scale could be $u_h(T)$ and if
  $u_h\thicksim T^{-1}$ we have the condition $r_c\ll T^{-1}$ which
  leads to $RT\ll \k^{-\frac{1}{6}}$}. For the confining
backgrounds a sensible constraint would be that the radial cut off
scale $u_k$ is larger that the thickness of the F-string $r_c\ll
u_k$. %and $r_c\ll T^{-1}$.

To compare length scales, we use \eq{temp1} write $r_c$ in terms of $R$ as
\be
\label{condition1a}
r_c=\prt{\frac{k T_s}{6 A R^6}}^\frac{1}{6} R:= \frac{3^7}{2^5}\k^\frac{1}{6} R~,
\ee
where the dimensionless parameter $\k$ has been defined as
\be
\label{kdef}
\k:=
\frac{2^5}{3^7}\frac{ k T_s}{AR^6}=B(\s)\frac{\sinh\a}{\cosh^5\a}~,\quad \mbox{with}\quad
B(\s)=\frac{|g_{00}|^3}{\pi^6 T^6 R^6}~,
\ee
and the  normalization of $\k$ is chosen to simplify its relation to $B$ above.  From (\ref{condition1a}) we
obtain:
 \be\la{kk1} r_c\ll R~\Rightarrow \k\ll
1~.
\ee
The constant $\k$ can also be related to the parameters $\l,~N$
and $\k$ as
\be\label{kk} \k=\frac{2^7}{3^6}~ \frac{k \sqrt{\l}}{N^2}~,
\ee
 where we have used the definition of $A$ given in \eq{tens} and
the fact that $16 \pi G_{10}=(2\pi l_s)^8g_s^2/(2\pi)$, $\l=4 \pi g_s N$
and $\Omega_7=\pi^4/3$. Notice that by considering the supergravity
limit, the condition \eq{kk1} implies that $k\ll N$.

From (\ref{Tloc}) see that as we move into the bulk
$T_{loc}$ becomes large as the redshift increases -- at some point
$T_{loc}$ will become so large that the probe approximation breaks down.
For some backgrounds, this constraint, which translates into how
deep the worldsheet can extend into the bulk, can be extracted from \eq{kdef}.  If we  assume that, for the
background of interest, the redshift factor increases as we go deeper
into the bulk and using the fact that $F(\a)=\sinh\a / \cosh^5\a$ has a
maximal value, one sees that for fixed $\k$, \eq{kdef} puts an upper
bound on the redshift allowed for the string.   Using
the maximal value of $F(\a)_{max}=16/\prt{25 \sqrt{5}}$ gives the
inequality
\be\label{crit1} \frac{\k}{B(\s)}\leq
\frac{2^4}{25\sqrt{5}}~\Rightarrow~ |g_{00}(u)|^3\geq
\frac{25\sqrt{5}}{2^4} ~\k \pi^6 R^6
T^6%~\simeq 3.49 \k \pi^6 R^6 T^6~.
\ee
Let $\s_c$ be the value of $\s$ which saturates the inequality
(\ref{crit1}) so that $\s\le \s_c$ (recall that we take the boundary
at $\s_b=0$).

Another constraint is generated by the requirement that the variation
of the local temperature should be small over the length of the string
probe:
\be
\frac{r_c T_{loc}'}{T}=-\frac{1}{2}\frac{r_c |g_{00}|'}{|g_{00}|^{3/2}}\ll 1~.
\ee
 This ensures that the string can be regarded as a probe locally
in the background.

In general, in order to satisfy the probe approximation limit,
especially for the conformal backgrounds, we should keep the turning
point of the string relatively far away from the horizon of the black hole of
 the background. However, in the confining theories these constrains are weaker or
even absent.

%From the $\k$ constrains we see that if needed, Taylor expansions on certain parameters could be done and that is what we do to the next section.

\subsection{Regularized free energy}
\label{sec:regF}

 We  eliminate the divergences in the free energy using the usual
subtraction of infinite bare quark masses. This translates to
subtracting the free energy of two infinite straight strings, $\cF_{\parallel}$, with shape
\be
x_0=\t,\quad u=\s~,
\ee which gives
\be
\label{Fpara}
\cF_{\parallel}=\frac{\sqrt{\l}k}{\pi R^2}\int_{\s_b}^{\s_c}
d\s\sqrt{f_u(u)} Y~,
\ee where $\s_c$ is the critical point saturating (\ref{crit1}) defined earlier. It follows that the total regularized free energy
for a Q\={Q} Wilson loop configuration is given by
\be\label{freg}
\cF_{reg}=\cF-\cF_{\parallel}=\frac{\sqrt{\l}k}{\pi
  R^2}\left(\int_{\s_b}^{\s_0} d\s\sqrt{f_u(u)} Y\left(\sqrt{1+
      f_{1u}(u) x'^2 }-1\right)-\int_{\s_0}^{\s_c} d\s\sqrt{f_u(u)}
  Y\right)~,
\ee
where the $x'$ is given by \eq{xp}. To express the free energy in
terms of the length, $L$, one should integrate \eq{l1} and
solve for $\s_0$. The expression $\s_0\prt{L}$ can then
be inserted into \eq{freg} to get the static
potential in the form $V\prt{L}$. This procedure is not usually doable
analytically without approximations so we find an expansion
for the static potential using $\k\sim k/N^2$ as an expansion parameter.

The last integral of equation \eq{freg} has $\s_c$ as a lower limit. However, close
to $\s_c$ the probe approximation for the string in certain
backgrounds, might break down. This is usually the case for the
conformal backgrounds, while for the confining backgrounds we mostly focus on here, we do not have this problem. For a possible resolution of the
issue in the conformal backgrounds, where we can take $\sqrt{f_u}\sim1/\s^2$, we can write energy from the
equation \eq{freg} as
\bea\nn
\cF_{reg}=&&\frac{\sqrt{\l}k}{\pi R^2}\bigg(\int_0^{\s_0} d\s\sqrt{f_u(u)} \prt{Y\left(\sqrt{1+ f_{1u}(u) x'^2 }\right)-1}-\\\la{ob}
&&\prt{\frac{1}{u_0}-\frac{1}{u_c}+\int_{0}^{\s_c} d\s\sqrt{f_u(u)} \prt{1-Y}}\bigg)~.
\eea
In the limit $\k \rightarrow 0$ the last term approaches zero and
 $\s_c\rightarrow \s_h$. Since, in this
case the infrared $\s_c$ dependence is eliminated,
we seem to  have a sensible  regularized energy for the conformal backgrounds\footnote{For more details on $\s_c$ and the treatment of the infrared cut-off of \eq{Fpara} see \cite{Grignani:2012iw}}. However, although the result is independent of the IR cutoff,
special care is needed since, the UV divergence
cancellation need to be equivalent to the Legendre transform method
\cite{Chu:2008xg,Drukker:1999zq}, and the contributions of the infinite strings free energy, $\cF_{\parallel}$,
might not be properly taken into account. However, in the confining
backgrounds we are mainly interested in, this is not a problem -- we
will see below that the usual  choice for
subtraction of the UV divergences is valid.

%An other sensible choice would be to perform the calculation with the current limits, and a critical upper boundary $\s_c$ for the integrals, up to where the probe approximation is valid. It turns out that the total energy of the Wilson loop in sensitive to the choice of the infinite string length.
\subsection{Analytic expressions for small $\k$ expansion}

The equation \eq{kdef} can be written as a quintic in
$\cosh^2\alpha$:
\begin{equation}
  \label{eq:quintic}
  \left(\frac{\kappa}{B}\right)^2X^5-X+1 =0 \qquad \mathrm{where}\quad X=\cosh^2\alpha~.
\end{equation}
For $(\kappa/B)^2<256/3156$, this has two real positive
solutions. Taking $\kappa\rightarrow0$ one of the solutions
corresponds to $\a\rightarrow 0$ and the other to $\a\rightarrow
\infty$. The extremal limit of the branes is obtained by sending the
local temperature to zero while at the same time $r_0\rightarrow 0$
and $\a\rightarrow \infty$ keeping the total charge fixed. This means
we should take the second root which may be expanded as
\be
\begin{split}
\a &\simeq
\mathop{\mathrm{arccosh}}\left(\sqrt{\frac{\sqrt{B(\s)}}{\sqrt{\k}}-\frac{1}{4} -\frac{5}{32}\frac{\sqrt{\k}}{\sqrt{B(\s)}}}\right)\\
&\simeq\frac{1}{4}\log\left(\frac{16 B(\s)}{\k}\right)-\frac{3\sqrt{\k}}{8\sqrt{B(\s)} }+\cO \prt{\frac{\k}{B(\s)}}~.
\end{split}
\ee
We may now expand the equation of motion \eq{xp} and subsequently the length $L$ without reference
to explicit metric components giving
\bea\label{ls}
&&\frac{L}{2}=\int_0^{\s_0} d\s\left(l_0(\s)+\sqrt{\k}~ l_1(\s)+\cO \prt{\k}\right),\where  \\
&&l_0(\s):=\sqrt{\frac{f_1(\s_0) }{f_{1u}(\s) (f_1(\s) -f_1(\s_0))}}~,\quad\mbox{and}\\
 &&l_1(\s):=l_0(\s) \frac{R^3\pi^3 T^3}{3 }\frac{\left(|g_{00}(\s_0)|^{3/2}-|g_{00}(\s)|^{3/2}\right)g_{11}(\s)}{\left(f_1(\s)-f_1(\s_0)\right)
|g_{00}(\s_0)|^{3/2}\sqrt{|g_{00}(\s)|}}~.
\eea
When the integrals can be done analytically the inversion of \eq{ls} may be possible. The free energy can be also  expanded in small $\k$ giving
\be
\label{E1234}
E=\int_0^{\s_0}d\s\prt{E_1+\sqrt{\k} E_2}-\int_{\s_0}^{\s_c}d\s
\prt{E_3+\sqrt{\k}E_4}+ \cO \prt{\k}~,
\ee
where the exact expressions are presented in the Appendix \ref{app:analytice}. Notice the second term
has a further implicit dependence on $\k$ due to the upper bound,
$\s_c$, of the integral.  For particular confining backgrounds this dependence might be avoided as we explain below.

\section{Blackfold approach in confining backgrounds}

In this section we apply our  results to confining theories. We
examine the relevant thermal excitations of the strings and obtain
modifications of the static potential related to the way that the
string tension is modified. It is known that for large separation, the
static potential has the linear term indicating confinement and higher
order corrections depending exponentially on $(-L)$, where $L$ is the length
\cite{Giataganas:2011nz,Giataganas:2011uy,Nunez:2009da}. We  analytically compute the potential
of heavy Q\={Q} pairs in the blackfold approach, using
the confining theory of the a temperature $AdS_3$ soliton background.
Corrections to the term linear in $L$ indicate a decrease of the
string tension. This is naturally expected as we elaborate below.

\subsection{Confining Theory in the low Temperature Phase}

The thermal soliton $AdS_3$ background in the low temperature phase
corresponds to a confining theory.  The background has
the form
\bea\la{metrica1}
&&ds^2=\frac{u^2}{R^2}\prt{-dt^2+dx_1^2+dx_2^2+f\prt{u}d\phi^2}+ \frac{R^2}{u^2 f\prt{u}}du^2~,\\
&&f(u)=1-\frac{u_k^4}{u^4}~.\label{eq:f}
\eea
%It consists of three spatial
%dimensions one of them compactified on $S^1$ with radius $\rho$.
The submanifold parametrized by the coordinates $u$ and $\phi$ has the
topology of a cigar.
The tip of the cigar is at $u=u_k$ which introduces a new scale.
To avoid a singularity at the tip, the following periodicity conditions must hold
\bea\la{period1}
&&\phi \sim \phi+ 2\pi \rho~,\quad\quad \rho:=
\frac{R^{2}}{u_{k}}~ .
\eea
The periodicity of the Euclidean time $t$ is associated with the
inverse temperature $1/T:=2 \pi u_h$ and can be freely chosen.  While the time circle never shrinks to zero, the
circle in the $\phi$ direction shrinks to zero at the tip . %This
%justifies that the background geometry in low temperature is similar
%to the geometry of \eq{metrica1}.

By exchanging the $\phi$ and $t$ circles we obtain a blackhole background with same the asymptotics, corresponding to the
deconfined phase:
\bea\la{metrics1}
&&ds^2=\frac{u^2}{R^2}\prt{-f\prt{u} dt^2+dx_1^2+dx_2^2+d\phi^2}+ \frac{R^2}{u^2 f\prt{u}}du^2~,\\
&&f(u)=1-\frac{u_h^4}{u^4}~.
\eea
 Here the time
circle shrinks to zero at the horizon, $u_h$, while the $\phi$ circle does not.
To avoid a singularity, we enforce a periodicity condition along the
time direction  similar to \eq{period1}. Conversely for this solution,
the $\phi$ circle  may have arbitrary periodicity.

By calculating the free energies of these two finite temperature
backgrounds, it can be seen that at low temperatures ($T\lesssim
1/2\pi \rho$) the confining background dominates while
the deconfined phase dominates at higher temperatures
\cite{Horowitz:1998ha,Aharony:2006da}.

Our main focus is on the low temperature confined background \eq{metrica1}  with the Euclidean
time direction compactified but we also study the high temperature deconfined phase \eq{metrics1}.

%The possibility of additional D8-\={D}8 flavors branes \cite{SS} to the backgrounds under study have been also considered, where the holographic properties of mesons have been studied extensively. It should be noticed however that the flavors are massless generating potential misconceptions with the string breaking scales and therefore for a rigorous approach we will be studying the background without any flavors degrees of freedom.

\subsection{Thermal Excitations in the Confining Phase Background}

Using the background \eqref{metrica1}, we calculate the separation of the Q\={Q} pairs and the regularized
free energy from \eq{l1} and \eq{freg} respectively. To get
analytic expressions we assume that the leading contributions come from the
region close to turning point -- this is where the string
almost totally lies when $u_0\rightarrow u_k$
\cite{Sonnenschein:1999if,Giataganas:2011nz}. We expand \eq{l1} in
$\k$   and treat each order separately  integrating order by order. As a final step to invert $u_0\prt{L}$ we
Taylor expand the result around $u_k$,  since the string world-sheet
position is taken very close the tip.

After some algebra we find the separation  of the pairs to be
\bea\label{ssl1}
&&L=\frac{R^{2}}{2 u_k}\prt{c_1-\log\left(\frac{u_0}{u_k}-1\right)} +\sqrt{\k} \frac{\pi^3 R^8 T^3}{8 u_k^{4}} \left(c_{2}+\log\left(\frac{u_0}{u_k}- 1\right)\right)~,
\eea
where we have omitted higher order terms of $\cO\prt{k}$.
The values of the constants $c_{1}<0$ and $c_{2}>0$ can be found but
play no role in the final result as we show below. Subleading constant
terms coming from the boundary have been neglected in
\eqref{ssl1}. Notice the appearance of a logarithmic term at both
$\cO\prt{\k^0}$ and $\cO\prt{\k^{1/2}}$. This hints that higher orders
of $\k$ contribute to the linear term as well as to the exponential
corrections. On top of that the sign difference between the logarithmic terms hints
that the contribution will not be additive which we verify below. We remark that the above formula for the dipole
distance $L$ is valid only when the length is large.

 Solving
for $u_0$ in \eq{ssl1} and expanding in $\k$ we obtain
\bea\label{u0invert}
u_0=u_k\prt{1+e^{c_{1}-\frac{2 L u_k}{R^2}}}+\sqrt{\k} e^{c_{1}-\frac{2 L u_k}{R^2}} \frac{\pi^3 R^4 T^3 }{ 4 u_k^2}\prt{\prt{c_{1}+c_{2}} R^2-2 L u_k}~.
\eea
The  term of  $\cO \prt{\sqrt{\k}}$ in \eqref{u0invert} has a similar form to  the zero-th order term with a sign difference and an  additional  term of the form $\sim L \exp\prt{- L}$.

The leading contributions to energy come from the region of the string around $u_k$ and are calculated from \eq{freg}. Therefore, we integrate  \eq{freg} focusing on the leading contributions around the turning point of the string  close to the tip of the geometry.

After some algebra  and taking into account carefully the contributions of the incomplete Elliptic integrals we find
\be
\cF_{reg} = \frac{\sqrt{\l}k}{\pi R^2}\prt{\frac{u_k}{4}
\prt{ c_{e1} -\log\left(\frac{u_0}{u_k}-1\right)}+ \sqrt{\k}\frac{\pi^3 R^{6}T^3}{48 u_k^{2}}\prt{c_{e2} +7 \log\left(\frac{u_0}{u_k}-1\right)}}~,
\ee
where $c_{e1}<0$ and $c_{e2}>0$ are constants that are determined but
have subleading effects. We observe that the order $\k^0$ and $\k^{1/2}$
 terms are of the same form albeit with different signs. To
express the energy in terms of the distance between  the pairs we use  \eq{u0invert} which gives
\be\label{energyff}
\cF_{reg}=\frac{\sqrt{\l}k}{2\pi R^2}\prt{C_1 u_k+\prt{\frac{u_k}{R}}^{2} L +\sqrt{\k}\frac{\pi^3 R^{4} T^3}{3 u_k^{2}}\left(C_2 R^2-
L u_k\right)}~,
\ee
where $C_1$ and $C_2$ are constants given by
\be
C_1=\frac{c_{e1}-c_1} {2}~,\qquad  C_2=\frac{1}{8} \prt{4 c_1-3 c_2+c_{e2}}~.
\ee
Using the final regularized energy expression we find that the string tension becomes
\be\la{ell1}
\cF_{linear}=\frac{\sqrt{\l}k}{2\pi R^2}\left(\prt{\frac{u_k}{R}}^{2} -\sqrt{\k}\frac{\pi^3 R^{4} T^3}{3 u_k}
 \right)L~.
\ee
Therefore using the blackfold approach, we find that the string tension of the tube between the heavy quark bound state is decreased as  temperature is increased. This
is in agreement with the expectations, since a bound state like
quarkonium dissociates more easily  with increasing temperature.

A comment on the additional term in the tension \eq{ell1} and the results of effective string theory models, is in order. For large strings at finite temperature the string tension has been calculated using effective string models and the Nambu-Goto action. The low temperature expansion of the string tension was found to be \cite{Pisarski:1982cn}
\be
\s\prt{T}=\s_0 - \frac{\prt{d-2} \pi}{6}T^2+\sum_{n\ge 3} a_n T^n~,
\ee
where $\s_0$ is the string tension in zero temperature. The second term is similar to the L\"{u}scher term of the static potential and does not depend on the gauge group of the theory considered.
Considering a version of open string-duality \cite{Luscher:2004ib}, it was shown that the cubic term is absent since  $a_3=0$.  Here we are bringing $k$-interacting strings to thermal equilibrium with the background and we calculate the string tension of the bound state. Unfortunately taking the limit of the single string is not smooth since the blackfold approach is not valid anymore. Therefore, we can not predict with certainty if a cubic term arises in the single string tension when a string is in thermal equilibrium with the theory.

Moreover, notice that the critical distance obtained by \eq{crit1} is
\be
u_c^3=\frac{5^{5/4}}{4}\pi ^{3}\sqrt{\k} R^{6} T^{3}
\ee
which does not impose any strict constraint on the system since is close to the boundary and can always be taken such that is less than the $u_k$.
Furthermore requiring that the width of the F-string, $r_c$ is much smaller  $u_k$  leads to $\k^\frac{1}{6}\ll u_k/R$ which does not constrain our calculations.

The analytical results obtained in this section are  supported below by  numerical computations done using the full system of equations .

\subsection{Numerical Analysis}

In this section we numerically compute the energy of the bound state
in terms of the distance between the pairs. We use the full equations
of motion, and compute the potential $V\prt{L}$ without any
approximation for several values of $\k$ including $\k=0$. Then we do a linear fit
of the form $V=\s_\k L+c_\k$, where $\s_\k$ is the
string tension and $c_\k$ is a constant, both depending $\k$.
\begin{figure*}[!ht]
\begin{minipage}[ht]{0.5\textwidth}
\begin{flushleft}
\centerline{\includegraphics[width=80mm]{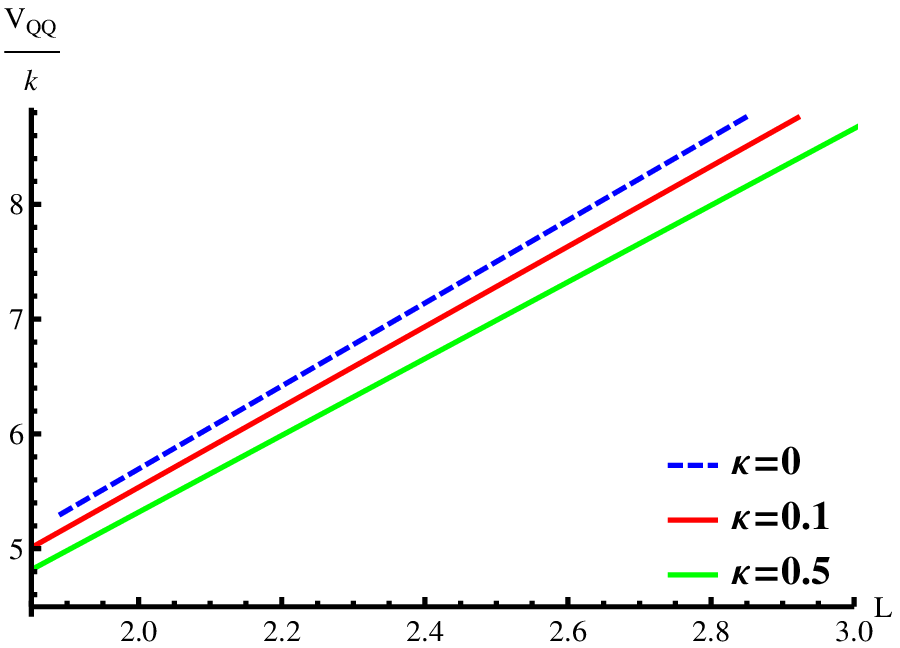}}
\caption{\small{The Energy of the bound state per number of pairs for $\k=0,~0.1~,0.5$, with
    temperature fixed, plotted for large $L$. The value $\k=0$
    corresponds to the usual Wilson loop Plateau problem with the
    Nambu-Goto action minimization, while for $\k\neq0$ the blackfold
    method is applied. Increasing  $\k$ decreases
     the  slope corresponding to a decrease of the string
    tension. This is in agreement with the analytical result obtained
    in \eq{energyff}. For the computation we have made the following choices:
    $R=1,~u_k=1.9 R,~T= 0.4/R$~. The energy is divided by
    $\sqrt{\l}/\prt{2\pi R^2}$ and the dimensionful quantity $R$ is
    used to make the $V_{QQ}$ and $L$ dimensionless. For $\k=0.1$ and $\k=0.5$, the critical
    distances where our method is valid, are  $u_{crit}\simeq 1.05
    R$ and $u_{crit}\simeq 1.38 R$
    respectively.
\vspace{0cm}}}\label{fig:l1}
\end{flushleft}
\end{minipage}
\hspace{0.3cm}
\begin{minipage}[ht]{0.5\textwidth}
\begin{flushleft}
\centerline{\includegraphics[width=80mm]{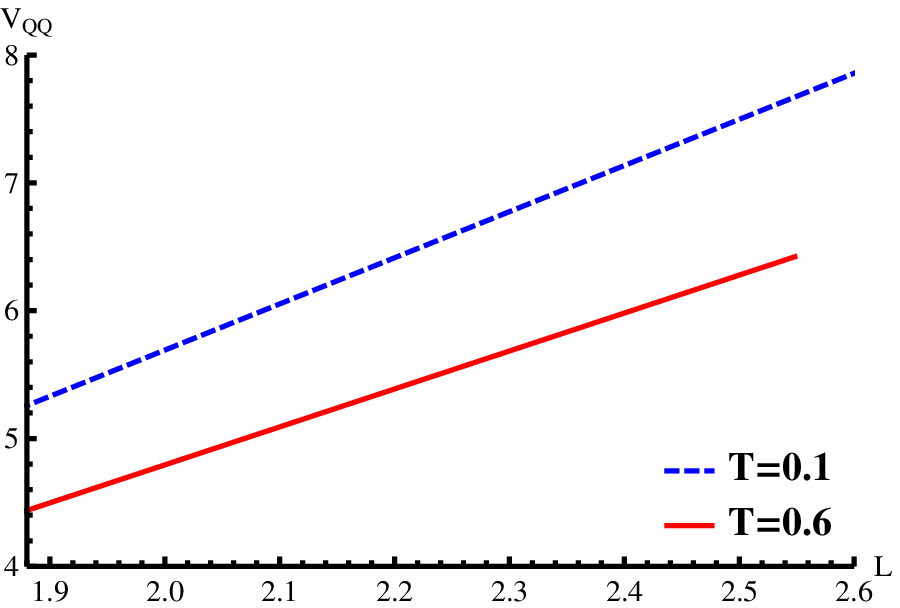}}
\caption{\small{The Energy of the bound state (in units of $R$) for $T=0.1$ and $T=0.6$ with $\k$ fixed and plotted for
    large values of $L$.  Increasing the temperature decreases the slope which
    translates to a decrease of the string
    tension in agreement with the analytical result
    \eq{energyff}. We have chosen $\k=0.2$, and the other parameters and normalizations to be the same as in Figure
    \ref{fig:l1}. For $T=0.1$ and $T=0.6$, the critical distances for our method
    to be valid are  $u_{crit}\simeq 0.3 R$ and $u_{crit}\simeq
    1.78 R$ respectively.
    \vspace{2.7cm}}}\label{fig:l2}
\end{flushleft}
\end{minipage}
\end{figure*}

We compare the plots and the fitted curves for different values of
$\k$ and  find  results supporting the analytic results of the previous
section. More specifically,  Figure \ref{fig:l1} shows  the $V_\k\prt{L}$ per number of strings $k$,
for $\k=0,~\k=0.1$ and $\k=0.5$  with $L$ large and temperature fixed. The $\k=0$ case corresponds
to the Nambu-Goto treatment, while for non-zero $\k$ the
blackfold terms appear. Increasing $\k$ decreases the slope and
therefore the string tension per string.  The linear fitting leads to
\bea\nn
\frac{V_k \prt{L}}{k}= \frac{\sqrt{\l}}{2\pi R^2}\prt{c_{\k}+\{3.61,3.50,3.35\} ~L}~,\quad\mbox{for}\quad \k=\{0,~0.1,~0.5\}~.
\eea
where $c_{\k}$ is a constant depending on $\k$ weakly and the number of strings $k$ is taken equal to 1
for $\k=0$. Note that in contrast to the potential per number of strings $V_k \prt{L}/k$, the total potential and the string tension
of the whole bound state is increased when the $\k$ is increased, since the number of strings $k$ is proportional to $\k$ from equation \eq{kk}.

Next we compute the potential of the bound state with a large size $L$ for two different values of the temperature
with $\k\neq0$ and fixed, and compare the obtained slopes. The outcome
can be seen in Figure \ref{fig:l2}. The linear fittings of the
potential gives
\bea\nn
&&V \prt{L}=\frac{\sqrt{\l}k}{2\pi R^2}\prt{c\prt{T}+\{3.60 , 2.98\}~L}~,\quad\mbox{for}\quad T=\{0.1,~0.6\}~,
\eea
where $c\prt{T}$ depends on $T$. The string tension of the bound state is decreased
with increasing temperature as our analytical results indicate \eq{ell1}. As mentioned in the previous section, this is expected since
the binding of  quarkonium becomes weaker at higher temperatures.

\subsection{Thermal Excitations in the Non-Confining phase}

In the high temperature phase, with background blackhole metric \eq{metrics1}, the dual gauge theory is no longer
confining. The  temperature can be found  from the periodicity of the Euclidean time circle and is given by $T=u_{h}/\prt{\pi
  R^{2}}$.

In  this case, Wilson loop calculations can in principle be done analytically.
The string configuration needed for the Wilson loop extends along the time, one spatial and the radial direction and is localized along the rest of the directions.
  Unfortunately, the scale introduced by the  black hole  leads to stronger constraints on  the validity of our method.

Applying our generic equations \eq{l1} and \eq{ob} to the background \eq{metrics1}, we obtain the deconfined results also obtained in \cite{Grignani:2012iw}
%\be
%\cF_{loop}=-\frac{\sqrt{\l}k}{L}\prt{\frac{4 \pi^2}{\Gamma\prt{\frac{1}{4}}^4}+\sqrt{\k} \frac{\Gamma\prt{\frac{1}{4}}^4}{96}\prt{L T}^3+\frac{3 \Gamma\prt{\frac{1}{4}}^4}{160}\prt{L T}^4 + \dots}
%\ee
\be\la{adsfin1}
\cF_{loop}=-\frac{\sqrt{\l}k}{L}\prt{\frac{4 \pi^2}{\Gamma\prt{\frac{1}{4}}^4}+c \frac{\sqrt{k} \l^{\frac{1}{4}}}{N} \frac{\Gamma\prt{\frac{1}{4}}^4}{96}\prt{L T}^3+\frac{3 \Gamma\prt{\frac{1}{4}}^4}{160}\prt{L T}^4 + \dots}~.
\ee
In the above formula we notice a remarkable similarity with some already known results. The $1/N$ correction happens to be, up to constants, the square root of the energy of the higher representation Wilson loops. For example
the circular Wilson loop in the $k$-th symmetric product
representation of $U\prt{N}$, corresponding to a D3 brane, is
found to have a term proportional to $k^3 \l^{3/2}/N^2$
\cite{Drukker:2005kx} which is the square of the $\cO\prt{k^{3/2}}$
term of \eq{adsfin1}. The higher representation Wilson loops   can be
thought as representing interacting Q\={Q} strings of $k$ different
pairs, and the interaction term in the energy always appears  as a multiple
of  $1/N^2$. In the blackfold approach we have at zero temperature $k$ non-interacting strings, and the nature of this term may be due to some indirect
interaction between the strings that is introduced when the string is
brought into thermal equilibrium with the background.

%In this background the constrains of the section \ref{sec:con} play a more direct role to the Wilson loop calculation. We present the results for two different regularization of the infinite strings. Once the infinite string is taken strictly for lengths that to ensure probe string approximation and the other is taken using the limit $\k\rightarrow 0$ where the
%two last terms of the \eq{ob} become $1/uh$ and $0$ respectively. In the first case we integrate the equations \eq{l1} and \eq{freg} for the length and the energy of the loop respectively, while for the second one we integrate the equations \eq{l1} and \eq{ob}. We present our results in Figures 2 and 3.

%For the non-conformal background as it is this one we observe that the results are so sensitive of the total energy with respect to length that different normalizations of the infinite strings lead to different qualitative behavior. The normalization with $\k\rightarrow 0$, leads to results in agreement with \cite{oberswl} where the same normalization was used.

\section{Conclusions}

We have developed  a generic formalism applicable to
any background (with diagonal metrics), which describes the string
world-sheet dual to a Wilson loop using the blackfold approach. Our
formulas are in terms of the metric elements. In the energy of
the bound state, the thermal excitations of the string due to the
finite temperature of the theory have been captured. This is along the
same lines as the string Nambu-Goto treatment where several
observables, including the energy of the heavy quark bound state, have been
expressed in terms of metric elements in generic formulas, applicable
to large classes of backgrounds.

We have applied our generic formalism to a confining theory, the AdS
soliton, and have found that the string tension is reduced due to the
thermal excitations with a negative term proportional to $c \sqrt{\k}
T^3/u_k$, where $c$ is a dimensionful constant that has been found. We
remark that this is the first time where AdS/CFT top-down techniques have been used to find a
temperature dependent observable related to heavy quarks in the confining phase of the theory. To capture the temperature dependence of
an observable in  a confining theory at low temperature through
AdS/CFT is a difficult task, due to the fact that the finite
temperature and the confining scale are introduced by a time and space
compactification respectively that are competing with each other, and only
one circle can shrink to zero at a time. This is why  the thermal
confining backgrounds, below the Hawking-Page transition have no black
hole in their horizon. So the only way one can see the
effect of the temperature in an observable is to introduce it in a
different way which is  what we have done here.

The effect of the temperature on the string tension, we have found,  has some interesting features. First of all, this is exactly what is expected to happen qualitatively to the string tension when the heavy bound state starts to heat up. This has been also confirmed by the string motivated
models and Monte Carlo lattice computations
\cite{Pisarski:1982cn,deForcrand:1984cz,Gao:1989kg,Bakry:2010zt}. We note that the additional terms depending on the temperature, are not
exactly the same as the ones that the string models predict.  In retrospect this is not too surprising given differences between the two
approaches and the bound states considered. Here we have $k$ number of strings in thermal equilibrium with the background and we calculate the string tension of the bound state, where the limit of the single string is not smooth since there the blackfold approach is not valid anymore. Therefore, we can not predict with certainty that the terms we have found arise also in the expression for the tension of a single string in thermal equilibrium with the background. Moreover, we expect that the appropriately modified methods of the effective string models, should produce analogous terms in the string tension in the context of AdS/CFT, making our approach in a sense complementary to those.

It is essential to point out that we interpret the static potential
that we have calculated as coming from a number of
Q\={Q} pairs in the fundamental representation which we consider to be  initially non-interacting at zero
temperature. Bringing the strings into thermal
equilibrium with the bath affects  the energy of the stack of
the strings considerably. The terms generated with the blackfold approach depend on the number $k$ of the strings in such a way that it can not be factored out as a common factor.
This implies that, even indirectly, there is an interaction between the
flux tubes of the different pairs while bringing them into thermal
equilibrium. This configuration is different from  the
bound state of the $k$-strings made by the interaction of the gluonic strings and represented in the
gravity dual  as $D3$ or $D5$ branes. We nevertheless found a
similarity in the mathematical expressions between the $k$-string energy term and the term generated by the blackfold term.

It would be very interesting to study the string tension's dependence on the temperature in larger class of confining theories and dimensions. This would give information on how the string tension is modified among the different theories in low temperature confining phase. Currently there are limitations of the blackfold approach application to theories with non-constant dilaton and there are not many confining theories where our calculations can be directly extended. Nevertheless, this is an interesting future direction.

Our approach has a wide range of further applications for finding the
qualitative properties of observables depending on the temperature in
confining theories. %This task without the use of the blackfold is
%difficult due to the fact that the dual gravitational backgrounds in the confining phase do not depend on the
%temperature.
Extending our work, it should be possible to take into
account the thermal excitation of the strings by applying the
blackfold approach and obtain finite temperature effects on other
observables.

\section*{Acknowledgements:}
We are thankful to J. Armas and R.Emparan,
for useful conversations, comments and correspondence. The research of D.G. is partly supported by a Fellowship of State Scholarships Foundation, through the funds of the ``operational programme education and lifelong learning''  by the European Social Fund (ESF) of National Strategic Reference Framework (NSRF) 2007-2013. The work of K.G. is supported in part by the National Research Foundation.

\startappendix

\Appendix{Semi-analytic expressions for the energy for small $\k$ expansion}\label{app:analytice}

In this appendix we present explicit expressions for the quantities in \eqref{E1234}:\nn
\bea\label{energies}\nn
E_1&=&\sqrt{f_u(u)}\left(\sqrt{\frac{f_1(u)}{f_1(u)-f_1(u_0)}}-1\right)\\\nn
E_2&=&-\pi^3 R^3 T^3 \frac{\sqrt{g_{uu}(u)}}{3 |g_{00}(u)| |g_{00}(u_0)|^2\left(f_1(u)-f_1(u_0)\right)}\Bigg(-f_1(u)|g_{00}(u_0)|^{2}
+f_1(u_0)|g_{00}(u_0)|^{2}\\\nn
&&+\sqrt{\frac{1}{f_1(u)-f_1(u_0)}}\Big(f_1(u)^{3/2}
|g_{00}(u_0)|^{2}-2f_1(u_0)\sqrt{f_{1}(u)} |g_{00}(u_0)|^{2}\\\nn
&&+f_1(u_0)|g_{00}(u)|^{3/2}\sqrt{|g_{00}(u_0)|f_1(u)}\Big)\Bigg)\\\nn
E_3&=&\sqrt{f_u(u)}~,\qquad\qquad E_4=-\frac{\pi^3 R^3 T^3 \sqrt{f_u(u)}}{3 |g_{00}(u)|^{3/2}}~.
\eea
Notice that the integral $\int_{\s_0}^{\s_c}d\s
\prt{E_3+\sqrt{\k}E_4}$ in most theories has implicit dependence on $\k$ from the
upper bound of the integral $\s_c$ which is obtained by saturating the inequality \eq{crit1}.
From the above expressions for  the energy and the expression \eq{ls} for the length we already
notice an initial dependence of the quantities in terms of the temperature as $T^3$.

\bibliographystyle{JHEP}
\bibliography{botany}
\end{document}